\documentstyle[12pt]{article}
\def\G{\Gamma}
\def\g{\gamma}
\def\p{\partial}
\def\c{\chi}
\def\l{\lambda}
\def\ba{\begin{array}}
\def\ea{\end{array}}
\def\hs{\hspace}
\def\cd{\cdot}

\def\ot{\otimes}
\def\w{\wedge}
\def\Lra{\Longrightarrow}
\def\sra{\rightarrow}
\def\O{\Omega}
\def\o{\Omega}
\def\ll{\label}
\def\re{\ref}
\def\trd{\bigtriangledown}
\def\bea{\begin{eqnarray}}
\def\eea{\end{eqnarray}}
\def\be{\begin{equation}}
\def\ee{\end{equation}}
\def\nn{\nonumber}

\begin{document}

\parskip 5pt plus 1pc
\parindent=20pt
\begin{flushright}
April, 1994
\end{flushright}
\vspace{5ex}
\centerline{\Large \bf Gravity and Discrete Symmetry }
\vspace{10ex}
\centerline{\large \sf Bin Chen$^{a}$,~~Takesi Saito$^{b}$,~~
and~~ Ke Wu$^{a}$}
\vspace{2.5ex}
\centerline{{\sf a. Inst. of Theor. Phys., Academia Sinica.
P.O.Box 2735, Beijing 100080, China}\footnote{The project is
supported in part by National Natural Science Foundation of
China, and "Climbing Up" Foundation of China.}}
\centerline{\sf b. Dept. of Phys., Kyoto Pref. Univ. of Medicine,
Taishogun, Kyoto 603, Japan.\\}

\begin{minipage}{5in}
\vspace{20ex}
\centerline{\large \sf Abstract}
\vspace{3ex}

{\it
\hskip15pt Using the differential calculus on discrete group, we
study the general relativity in the space-time which is the
product of a four dimensional manifold by a two-point space. We
generalize the usual concept of frame and connection in our
space-time, and from the generalized torsion free condition we
obtain an action of a scalar field coupled to Einstein gravity,
which may be related to the Jordan-Brans-Dicke theory.}

\end{minipage}

\vspace{4ex}

\newpage

\section{Introduction}

In order to overcome some essential problems in elementary particle
models, the concept of non-commutative geometry has been introduced
in physics in recent years. It was Connes who first put forward a
picture in which his non-commutative geometry was used to construct
particle physical model \cite{Con1},\cite{Con2}.
Then, many others expounded his thoughts in other methods which are
not so abstruse in mathematics. There are so many works on this
problem that we can only give out uncompleted references, for
example see \cite{Kas,Co,Si1,Din,Mor,Na}. Some improvements have
been made in this direction. Especially, taking the Higgs field as
the gauge field on the internal discrete group, we can reconstruct
the standard model for electroweak-strong interactions in an elegant
and consistent way \cite{Si1},\cite{Din},\cite{Mor}.
Although we still don't know how to quantize these
reconstructed models, we have a possible explanation of geometrical
origin of the Higgs field and the symmetry breaking mechanisms,
which had puzzled physicists for many years.

Another interesting use of non-commutative geometry is in gravity.
Due to our shortage of knowledge on how to describe the space-time
at tiny distance, how to describe gravity at distance of order
Plank-length is a problem for us. It seems that we should modify
our classical geometrical concepts. So some persons have studied the
gravity using the non-commutative geometrical method.
First, Chameseddine et al. \cite{Che} study the general relativity
in the framework of Connes' non-commutative differential geometry.
Based on a space-time which is the product of four dimensional
manifold by a two-point space, they obtain a model of a scalar
field coupled to Einstein gravity. And they interpret this scalar
field as describing the distance between the two points in the
internal space. Very recently, some other attempts have been
made in this area \cite{Si2}, \cite{Ka}.

In this paper, by extending the formalism of \cite{Din}, we
investigate the relation between the discrete symmetry and
gravity. We also take the space-time as the product of four
dimensional manifold by a two-point space. In this space-time,
we generalize the usual concepts of frame and connection.
Beyond the ordinary frame and connection components on four
dimensional space-time, we introduce an extra component which
associates with discreate symmetry. After imposing our generalized
torsion free condition, we get some relations among connections.
We then calculate the curvature tensor, and obtain a gravity
action which describes a scalar field coupled to Einstein
gravity. This scalar field is closely related to the discrete
symmetry, and we would like to take it as the Brans-Dicke field
\cite{Bra}. We believe our formalism is more mathematically
concise and elegant than others.

This paper si scheduled as
follows: In \S2, we review the differential calculus on
discrete groups briefly. In \S3, we give out our generalization
of frame, connection, and the consequence of generalized
torsion free condition. In \S4, we calculate the curvature
tensor and give out our gravity action.  The final section
is devoted to some conclusions and remarks.

\section{Introduction to differential calculus on $M\ot Z_2$}

The details of differential calculus on $M\ot Z_2$ can be found
in \cite{Si1}, \cite{Din}. In this section we give out some
basic and necessary information associated with our following
discussion.

Let $\cal A$ be the algebra of complex valued
function on $M\ot Z_2$ and let $d_M,d_{Z_2}$ be the external
derivative on the differential algebra on $M,Z_2$ respectively.
The external derivative $d_{Z_2}$ acts on $\cal A$ as follows:
\be
d_{Z_2}f={\p}_{Z_2}f{\c}=(f-R(f))\c, \hs{3ex} \forall f\in
{\cal A}. \ll{2.1}\ee
Here $\c$ is a given one-form of $Z_2$ and $R$ is the right
action on $\cal A$:
\be (Rf)(x,e)=f(x,r),\hs{3ex} (Rf)(x,r)=f(x,e), \hs{3ex}
e,r\in Z_2. \ll{2.2}\ee
We denote the one form space as ${\O}^1$ and its basis as
($dx^m, \c$).
Then we can give out the external derivative on $\cal A$:
\be df={\p}_{\mu}f{\cd}dx^{\mu}+{\p}_{Z_2}f\cd \c. \ll{2.3}\ee

In order to satisfy the properties of the ordinary external derivative
operator, we should add two conditions on $d_{Z_2}$:
\bea
\c f=&(Rf)\c, \ll{2.4}\\
d_{Z_2}\c=&-2\c \ot \c, \ll{2.5} \eea
and the nilpotency of $d$ requires that
\bea
dx^{\mu}\ot dx^{\nu}&=-dx^{\nu}\ot dx^{\mu}, \ll{2.6}\\
 dx^{\mu}\ot \c&=-\c \ot dx^{\mu}.  \ll{2.7} \eea

Let us define the metric $g$ as a form on the left module of
one-forms valued in the algebra $\cal A$ and bilinear over
algebra $\cal A$,
\be
g: {\O}^1\ot {\O}^1 \sra {\cal A} \ll{2.8}.\ee
Imposing the middle-linearity and hermity conditions on metric,
we obtain
the following results:
\be
<dx^{\mu}, dx^{\nu}>=g^{\mu \nu}, \hs{3ex}
<dx^{\mu}, \c> =<\c, dx^{\mu}>=0, \hs{3ex}
<\c, \c>={\eta}, \ll{2.9}\ee
where $\eta$ is a real parameter. Finally
we introduce the Haar integral as a complex valued linear
functional on
${\cal A}$ that remains invariant under the action of $R$
\be
{\int}_{Z_2}f=\frac{1}{2}(f(x,e)+f(x,r)). \ll{2.10}\ee.

\section{Generalized Connection and Torsion free condition}

Einstein's general relativity is based on the four-dimensional
space-time which is not suited to describe the gravity at small
distance. Now we take our space-time to be the product of
four-dimensional manifold by a two-point space, i.e. $M^4 \ot
Z_2$. In this space-time, the usual concept of frame and
connection should be generalized. Our generalization are as follow:\\
Frame
\be
 E^A=\left\{\ba{ll}
E^a=&e^a_mdx^m\hs{3ex} a,m=1,2,3,4\\
E^5=&{\l}\c
 \ll{3.1} \ea \right.\ee
Connection
\be
 {\O}^{AB}=\left\{\ba{l}
{\G}^{AB}_mdx^m\\
{\G}^{AB}_{\cd}\c \ea \right. \hs{3ex}
{\G}^{AB}_m=-{\G}^{BA}_m,\hs{3ex} {\G}^{55}_m=0 \ll{3.2}
 \ee
Here $A=a,5$ and $B=b,5$. $\l$ is a real scalar field. We must
note that our notation is a little
different from the usual one, here is the correspondence between
two notations
\be
{\G} \sra {\o} ,\hs{3ex}  {\g} \sra {\G} \ee
i.e. $\G$ to be spin connection and $\g$ to be affine connection.

We assume that the torsion free condition should be maintain,
i.e.$T^A=0$. Using
the generalized frame and connection, we calculate the torsion as:
\bea
T^A&=&dE^A+{\G}^{AB}_mdx^m{\ot}E^B+{\G}^{AB}_{\cd}\c{\ot}E^B,
\ll{3.3}\\
T^a&=&dE^a+{\G}^{ab}_mdx^m{\ot}E^b+{\G}^{a5}_mdx^m{\ot}E^5
+{\G}^{ab}_{\cd}\c{\ot}E^b
      +{\G}^{a5}_{\cd}\c{\ot}E^5 \nn\\
   &=&(d_x+d_{z_2})E^a+{\G}^{ab}_mdx^m{\w}e^b_ndx^n \nn\\
   & &+{\G}^{a5}_mdx^m{\ot}{\l}\c+{\G}^{ab}_{\cd}\c{\ot}e^b_ndx^n
+{\G}^{a5}_{\cd}\c{\ot}{\l}\c \nn\\
   &=&{\p}_me^a_ndx^m{\w}dx^n+[e^a_n-R(e^a_n)-\c{\ot} dx^n
+{\G}^{ab}_me^b_ndx^m{\w}dx^n  \nn\\
   & & +{\G}^{a5}_m{\l}dx^m{\ot}\c+{\G}^{ab}_{\cd}R(e^b_n)\c{\ot}dx^n
+{\G}^{a5}_{\cd}R({\l})\c{\ot}\c. \ll{3.4}
\eea
{}From the torsion free condition, we get the usual torsion free
condition
\be
({\p}_me^a_n+{\G}^{ab}_me^b_n)dx^m{\w}dx^n=0 \ll{3.5}\ee
and
\be
[e^a_n-R(e^a_n)-{\G}^{a5}_n{\l}+{\G}^{ab}_{\cd}R(e^b_n)-\c{\ot}dx^n=0.
\ll{3.6}\ee
For simplicity, we consider that the frame and connection on two
sheets be the
same, i.e.
\be
e^a_n=R(e^a_n), \hs{3ex} {\G}^{A}_{BC}=R({\G}^{A}_{BC}), \hs{3ex}
\l=R(\l),
\ll{3.7}\ee
then we find
\be
{\G}^{a5}_n{\l}={\G}^{ab}_{\cd}e^b_n. \ll{3.8}\ee
The other result from the torsion free condition is
\be
{\G}^{a5}_{\cd}R({\l})\c{\ot}\c=0 \Lra {\G}^{a5}_{\cd}=0.  \ll{3.9}\ee

Another component of torsion is
\bea
T^5&=&dE^5+{\G}^{5B}_mdx^mE^B+{\G}^{5B}_{\cd}\c{\ot}E^B \nn\\
   &=&d({\l}\c)+{\G}^{5b}_mdx^m{\ot}E^b+{\G}^{55}_mdx^m{\ot}{\l}\c
   +{\G}^{5b}_{\cd} \c{\ot}E^b
        +{\G}^{55}_{\cd}\c{\ot}{\l}\c \nn\\
   &=&{\p}_m{\l}dx^m{\ot}\c+({\l}-R({\l}))\c{\ot}\c-2{\l}\c{\ot}\c
+{\G}^{5b}_mdx^m{\w}e^b_ndx^n \nn\\
   & & +{\G}^{55}_{m}dx^m{\ot}{\l}\c+{\G}^{5b}_{\cd}\c{\ot}e^b_ndx^n
   +{\G}^{55}_{
   \cd}
\c{\ot}{\l}\c \nn\\
   &=&{\p}_m{\l}dx^m{\ot}\c+(-{\l}-R({\l}))\c{\ot}\c
+{\G}^{5b}_me^b_ndx^m{\w}dx^n \nn\\
   & & +{\G}^{55}_m{\l}dx^m{\ot}\c+{\G}^{5b}_{\cd}R(e^b_n)\c{\ot}dx^n
   +{\G}^{55} _{\cd}
R({\l})\c{\ot}\c. \ll{3.10}\eea
Also from the torsion free condition, we have
\bea
(1)& {\G}^{5b}_me^b_ndx^m{\w}dx^n=0 \Lra {\G}^{5b}_m=0 \nn\\
(2)& [{\p}_m{\l}+{\G}^{55}_m{\l}-{\G}^{5b}_{\cd}R(e^b_m)-dx^m{\ot}\c
=0 \nn\\
   &\Lra {\p}_m{\l}={\G}^{5b}_{\cd}e^b_m   \nn\\
(3)& [-{\l}-R({\l})+{\G}^{55}_{\cd}R({\l})-\c{\ot}\c=0 \nn\\
if &{\l}=R({\l}) \Lra {\G}^{55}_{\cd}=2, \nn\\
(if &{\l}=-R({\l}) \Lra {\G}^{55}_{\cd}=0).
\eea
In summary, from the torsion free condition we can get
\bea
{\G}^{a5}_{\cd}=0,& \ll{3.11a}\\
{\G}^{5b}_m=0&Lra {\G}^{b5}_m=0, \ll{3.11b}\\
{\p}_m{\l}={\G}^{5b}_{\cd}e^b_m,& \ll{3.11c}\\
{\G}^{55}_{\cd}=2 , (0).& \ll{3.11d}
\eea
The above results are very important and can be
used to calculate the curvature tensor and gravity action
in the next section.

\section{Curvature tensor and Gravity action}

In $\S$3 we generalized the frame and connection in $M^4 \ot Z_2$
and from the generalized torsion free condition we get some
important relations. Further in this section we use the
relations to calculate the curvature tensor in our space-time.
Finally we obtain the gravity action which describes a scalar
field coupled to the Einstein gravity.

{}From the frame and connection given above, we reach the
curvature tensor
\bea
R^{AB}&=&d\O^{AB}+\O^{AC}{\ot}\O^{CB} \nn\\
   &=&d({\G}^{AB}_mdx^m+{\G}^{AB}_m{\cd}\c)+({\G}^{AC}_mdx^m
   +{\G}^{AC}_{\cd}\c){\ot}
({\G}^{CB}_ndx^n+{\G}^{CB}_{\cd}\c)     \nn\\
   &=&{\p}_n{\G}^{AB}_mdx^n{\w}dx^m+{\G}^{AC}_m{\G}^{CB}_ndx^m
   {\w}dx^n \nn\\
   & & +d_{z_2}{\G}^{AB}_mdx^m+d_x{\G}^{AB}_{\cd}\c+{\G}
   ^{AC}_{\cd}\c{\ot}{\G}^{CB}_ndx^n
+{\G}^{AC}_mdx^m{\ot} {\G}^{CB}_{\cd}\c \nn\\
   & &+d_{z_2}{\G}^{AB}_{\cd}\c+{\G}^{AB}_{\cd}d\c
+{\G}^{AC}_{\cd}\c{\ot}{\G}^{CB}_{\cd}\c \nn\\
   &=&({\p}_m{\G}^{AB}_n+{\G}^{AC}_m{\G}^{CB}_n)dx^m{\w}dx^n
   +[{\G}^{AB}_m-R({\G}^{AB}_m)+{\G}^{AC}_{\cd}R({\G}^{CB}_m)-
   \c{\ot}dx^m \nn\\
   & &+({\p}_m{\G}^{AB}_{\cd}+{\G}^{AC}_m{\G}^{CB}_{\cd})dx^m{\ot}\c
    +[-{\G}^{AB}_{\cd}-R({\G}^{AB}_{\cd})+{\G}^{AC}_{\cd}R({\G}
    ^{CB}_{\cd})-
\c{\ot}\c. \ll{4.1}
\eea
The gravity action density can be calculated in usual way:
\bea
{\cal I}&=&<R^{AB},\hs{3ex} E^A{\ot}E^B> \nn\\
 &=&<R^{ab}_{mn}dx^m{\w}dx^n,E^a{\w}E^b>
+<(R^{5b}_{{\cd}m}-R^{5b}_{m\cd})\c{\ot}dx^m,E^5{\ot}E^b> \nn\\
 & &+<(R^{a5}_{m\cd}-R^{a5}_{{\cd}m})dx^m{\ot}\c,E^a{\ot}E^5>
+<R^{55}_{\cd\cd}\c{\ot}\c,E^5{\ot}E^5>, \ll{4.2}
\eea
where from (\re{4.1}) we find
\bea
R^{ab}_{mn}&=&{\p}_m{\G}^{ab}_n+{\G}^{ac}_m{\G}^{cb}_n \ll{4.3a}\\
R^{5b}_{{\cd}m}&=&{\G}^{5c}_{\cd}R({\G}^{cb}_m)={\G}^{5c}_{\cd}{\G}
^{cb}_m \ll{4.3b}\\
R^{5b}_{m\cd}&=&{\p}_m{\G}^{5b}_{\cd}+{\G}^{5c}_m{\G}^{cb}_{\cd}=
{\p}_m{\G}^{5b}_{\cd} \ll{4.3c}\\
R^{a5}_{m\cd}&=&{\p}_m{\G}^{a5}_{\cd}+{\G}^{ac}_m{\G}^{c5}_{\cd}=0
\ll{4.3d}\\
R^{a5}_{{\cd}m}&=&{\G}^{ac}_{\cd}R({\G}^{c5}_m)=0 \ll{4.3e}\\
R^{55}_{\cd\cd}&=&-{\G}^{55}_{\cd}-R({\G}^{55}_{\cd})+
{\G}^{5C}_{\cd}{\G}^{C5}_{\cd} \nn\\
 &&=(-2+{\G}^{55}_{\cd}){\G}^{55}_{\cd}. \ll{4.3f}
\eea
Finally, we obtain
\bea
{\cal I}&=&({\p}_m{\G}^{ab}_n+{\G}^{ac}_m{\G}^{cb}_n)
<dx^m{\w}dx^n,dx^p{\w}dx^q>e^a_pe^b_q \nn\\
 & &+({\G}^{5c}_{\cd}{\G}^{cb}_m-{\p}_m{\G}^{5b}_{\cd})
<\c{\ot}dx^m, {\l}\c{\ot}e^b_pdx^p> \nn\\
 & &+(-2+{\G}^{55}_{\cd}){\G}^{55}_{\cd}<\c{\ot}\c,{\l}\c{\ot}{\l}\c>.
 \ll{4.4} \eea
The first term in (\re{4.4}) is nothing but the usual curvature. But
the second term is not so obvious and needs some skills. From the
equation for frame
\be
{\trd}_me^b_p={\p}_me^b_p+{\G}^{bd}_me^d_p-{\g}^n_{mp}e^b_n=0,
\ll{4.5}\ee
we get
\bea
\lefteqn{
{\G}^{5b}_{\cd}{\G}^{bd}_{m}e^d_p-{\p}_m{\G}^{5b}_{\cd}e^b_p} \nn\\
&={\G}^{5b}_{\cd}{\G}^{bd}_{m}e^d_p-{\p}_m({\G}^{5b}_{\cd}e^b_p)
+{\G}^{5b}_{\cd}
{\p}_me^b_p \nn\\
&={\G}^{5b}_{\cd}({\p}_me^b_p+{\G}^{bd}_me^d_p)-{\p}_m({\p}_p{\l})
 \nn\\
&={\G}^{5b}_{\cd}{\g}^n_{mp}e^b_n-{\p}_m({\p}_p\l) \nn\\
&={\p}_n\l{\g}^n_{mp}-{\p}_m({\p}_p\l), \ll{4.6}
\eea
where (\re{3.11c}) has been used. The last equation is just the
covariant derivative
\be
{\trd}_mV_p={\p}_mV_p-{\g}^n_{mp}V_n, \hs{4ex} V_p={\p}_p\l,
\ll{4.7}\ee
so the second term in (\re{4.4}) becomes
\be
-{\trd}_m({\p}_p\l)\l g^{mp}\eta.  \ll{4.8}\ee
The third term in (\re{4.4}) is more simple if we use (\re{3.11d}),
we can easily find
that it vanishes. Since the frame and connection on two sheets are
the same, and we take $\l$ to be a real scalar field, the Haar
integral of ${\cal I}$ makes no difference.
At last we obtain the gravity action
\bea
I&\sim& R-{\trd}_m({\p}_p\l){\l}g^{mp}\eta \nn\\
&\sim& R+{\p}_p{\l}{\p}_m{\l}g^{mp}\eta. \ll{4.9}
\eea

\section{Conclusions and Remarks}

In this paper, we consider the gravity action in $M^4 \ot Z_2$.
Our formalism is based on differential calculus on discrete
groups. Analogue to usual construction in ordinary
four-dimensional space-time, we generalize the frame and
connection in our space-time.
Have considering the generalized torsion free condition,
we calculate the curvature tensor and then obtain our
final results. Our gravity action is very like the one
in \cite{Che}, but you can easily find that our formalism
is more concise and elegant than others \cite{Che},\cite{Si2},
\cite{Ka}, so it is easier to accept.

Finally, we want to give a few comments on the meaning of
the scalar field in our action. It is obvious that it must
be related to the discrete symmetry. In \cite {Che}, it was
interpreted as describing the distance between the two points
in the internal space. That is true. In our opinion we would
like to take it as a Brans-Dicke field. So, we obtain a
Jordan-Brans-Dicke theory. This is a very interesting
consequence.

\end{document}